\documentclass[english,12pt,aps,prd,a4paper,preprintnumbers,floatfix,nofootinbib,showpacs,superscriptaddress, notitlepage]{revtex4-1} 
 \pdfoutput=1
\usepackage[usenames,dvipsnames]{color}  
\usepackage{graphicx}
\usepackage{caption}
\usepackage{subcaption}
\usepackage{amsmath}
\usepackage{amssymb}
\usepackage[colorlinks=true,citecolor=darkred,urlcolor=darkred, pdfborder={0 0 0}]{hyperref}
\usepackage[normalem]{ulem}

\usepackage[T1]{fontenc}
\usepackage[latin9]{inputenc}
\usepackage{array}
\usepackage{booktabs}
\usepackage{mathrsfs}
\usepackage{multirow}
\usepackage{tabularx}

\definecolor{darkred}{rgb}{0.6,0,0}

\definecolor{linkcolor}{rgb}{0,0,0.5}

\def\gsim{\raise0.3ex\hbox{$\;>$\kern-0.75em\raise-1.1ex\hbox{$\sim\;$}}}
\def\lsim{\raise0.3ex\hbox{$\;<$\kern-0.75em\raise-1.1ex\hbox{$\sim\;$}}}

\def\beqn#1{\begin{equation}\label{#1}}
\def\eeqn{\end{equation}}

\def\beqa#1{\begin{eqnarray}\label{#1}}
\def\eeqa{\end{eqnarray}}

\def\Z2{$\mathcal{Z_2}$}

\newcommand {\ignore}[1]{}

\newcommand{\sm}{{Standard Model }}

\def\321{$\mathrm{SU(3) \otimes SU(2) \otimes U(1)}$ }
\newcommand {\black} {\color{black}}

\newcommand{\AddrAHEP}{%
  AHEP Group, Institut de F\'{i}sica Corpuscular --
  CSIC/Universitat de Val\`{e}ncia \\ 
 C/ Catedr\'atico Jos\'e Beltr\'an, 2 E-46980 Paterna, Spain}

\newcommand{\AddrFisteo}{%
Departament de F\'{i}sica Te\'orica, Universitat de Val\`encia, Burjassot 46100, Spain}

\newcommand{\AddrMiranda}{%
Departamento de F\'{\i}sica, Centro de Investigaci\'on
  y de Estudios Avanzados del IPN,\\ Apartado Postal 14-740 07000 Mexico,
  Distrito Federal, Mexico}
  
  \newcommand{\AddrIoannina}{%
Division of Theoretical Physics, University of  Ioannina, GR 45110 Ioannina, Greece}

\begin{document}

\bibliographystyle{unsrt}   

\title{\boldmath \color{BrickRed} XENON1T signal from transition neutrino magnetic moments }

\author{O.~G.~Miranda}\email{omr@fis.cinvestav.mx}\affiliation{\AddrMiranda}
\author{D.~K.~Papoulias}\email{d.papoulias@uoi.gr}\affiliation{\AddrIoannina}
\author{M.~T\'ortola}\email{mariam@ific.uv.es}\affiliation{\AddrFisteo}\affiliation{\AddrAHEP}
\author{J.~W.~F.~Valle}\email{valle@ific.uv.es}\affiliation{\AddrAHEP}

\begin{abstract}

  The recent puzzling results of the XENON1T collaboration at few keV electronic recoils could be due to the scattering of solar neutrinos endowed with finite Majorana transition magnetic moments (TMMs).
Within such general formalism, we find that the observed excess in the XENON1T data agrees well with this interpretation.
The required TMM strengths lie within the limits set by current experiments, such as Borexino, specially when one takes into account
a possible tritium contamination.


\end{abstract}

\maketitle

\section{Introduction}

Recently, the XENON1T collaboration has released very puzzling results at low recoils~\cite{Aprile:2020tmw}, which do not seem to fit with \sm (SM) expectations.
The data shows an excess over background, particularly pronounced in the low-energy tail around few keV.  
The XENON1T \emph{anomaly} prompted already many possible explanations see, e.g.,~\cite{AristizabalSierra:2020edu,Boehm:2020ltd, Lindner:2020kko,DiLuzio:2020jjp, Gao:2020wer, McKeen:2020vpf, Bally:2020yid, Khan:2020vaf}.
Given the experimental parameters, exposure, detection efficiency and energy resolution, one expects some sensitivity to solar neutrino backgrounds, especially from $pp$ neutrinos.
Here we investigate whether these findings could be indicative of the presence of new physics in the neutrino sector, beyond the simplest neutrino oscillation expectations.
Indeed, if neutrinos have finite transition neutrino magnetic moments~\cite{Schechter:1981hw}, there is a new component which adds to the electroweak neutral and charged current neutrino-electron
interaction cross section expected in the SM, given that this could dominate the scattering process at low recoil energies~\cite{Vogel:1989iv,Giunti:2014ixa}~\footnote{For their impact to astrophysics see, e.g. Ref.~\cite{Akhmedov:2003fu}.}.
We find that the excess seen in the XENON1T data is consistent with this interpretation.
The required magnitudes of neutrino transition magnetic moments (TMMs) is in agreement with known restrictions from experiments like Borexino, GEMMA or TEXONO, specially when the tritium component of the background is taken into account.
At the moment, since the latter is not fully understood, one can not jump to any big conclusions. It is, nevertheless, amusing to raise the possibility.
We also comment on whether the required parameter values could also be probed by astrophysics.
In contrast to other possible explanations given so far as to the possible cause of the low-energy event excess, ours does not suffer from severe tension with astrophysics.

\section{Formalism}
\label{sec:formalism}

In order to simulate the background due to the scattering of solar neutrinos on electrons at the XENON1T experiment, we take the differential event rate in terms of the reconstructed recoil energy, $T_{rec}$, as
\begin{equation}
\left[\frac{dN}{dT_{rec}}\right]_\mathrm{SM} = \varepsilon(T_{rec})\, \mathcal{E} \sum_{x} \int_{T_{e}^{\mathrm{min}}}^{T_{e}^{\mathrm{max}}} \int_{E_\nu^\mathrm{min}}^{E_\nu^\mathrm{max}} \, \frac{d \phi_x}{dE_\nu} \, 
\left[\frac{d\sigma_{\nu} (E_\nu, T_e)}{dT_e}\right]_\mathrm{SM} \mathcal{G}(T_{rec},T_e)\, d E_\nu \, dT_e   ,
\label{eq:dNdT_SM}
\end{equation}
where the index $x$ runs over all the solar neutrino components, $\phi_x$, of which the most relevant for the sensitivity range of
XENON1T are the continuous $pp$ flux and the monochromatic $^{7}\mathrm{Be}$ 861~keV line.  Here, $E_\nu$ is the neutrino energy,
$T_e$ is the true electron recoil energy, $\mathcal{E}$ denotes the number of electron targets contained in the 1042~kg detector multiplied by the exposure time of 226.9 days corresponding to a total 0.65~ton.yr
exposure, and $\varepsilon(T_{rec})$ is a detector efficiency factor. In Eq.~(\ref{eq:dNdT_SM}), the finite energy resolution of the detector is also taken into account by applying the smearing function $\mathcal{G}(T_{rec},T_e)$, approximated by a normalized Gaussian function with $\sigma/T_{rec}
  = (31.71/\sqrt{T_{rec} \, \mathrm{[keV]}} + 0.15) \%$~\cite{Aprile:2020yad}.
The SM differential cross section includes the contribution from all neutrino flavours as
\begin{equation}
\left[\frac{d\sigma_{\nu} (E_\nu, T_e)}{dT_e}\right]_{\mathrm{SM}}  = \frac{d \sigma_{\nu_e}}{dT_e} P_{ee}(E_\nu) + \frac{d \sigma_{\nu_{\mu,\tau} }}{dT_e} [1-P_{ee}(E_\nu) ]\, ,
\end{equation}
where  $P_{ee} (E_\nu)$ is the average survival probability for solar neutrinos reaching the detector.
The SM $\nu_e - e^-$ scattering cross-section receives contributions from both neutral current (NC) and charged current (CC) interactions, and is given by 
\begin{equation}
\left(\frac{d \sigma_{\nu_e}}{dT_e}\right)_{\mathrm{SM}} = \frac{2G_F^2 m_e}{\pi} \left[(g_L+1)^2 + g_R^2 \left(1 - \frac{T_e}{E_\nu} \right)^2 -(g_L+1) g_R \frac{m_e T_e}{E_\nu^2}\right] \, ,
\label{eq:xsec_SM1}
\end{equation}
with $G_F$ the Fermi constant and $m_e$ the electron mass. On the other hand, only NC interactions are involved in $\nu_\mu - e^-$
and $\nu_\tau - e^-$ scattering. The corresponding cross section reads
\begin{equation}
\left(\frac{d \sigma_{\nu_{\mu,\tau}}}{dT_e}\right)_{\mathrm{SM}} = \frac{2 G_F^2 m_e}{\pi} \left[g_L^2 +  g_R^2 \left(1 - \frac{T_e}{E_\nu} \right)^2 -g_L g_R \frac{m_e T_e}{E_\nu^2}\right] \, .
\label{eq:xsec_SM2}
\end{equation}
Here, the SM couplings $g_{L,R}$ are expressed in terms of the electroweak mixing angle parameter, $\sin^2 \theta_W$, as 
\begin{equation}
\begin{aligned}
g_L=& -1/2 +  \sin^2 \theta_W \,, \\
g_R=& \sin^2 \theta_W \, .
\end{aligned}
\end{equation}

In addition to the SM contribution, one can consider non-trivial electromagnetic (EM) neutrino interactions. These can be encoded in an effective neutrino magnetic moment  $\mu_{\nu,\mathrm{eff}}$, and are important at low recoil energies.
The presence of $\mu_{\nu,\mathrm{eff}}$ adds incoherently (due to helicity flip) to the SM differential cross section  in Eq.~(\ref{eq:dNdT_SM}) an electromagnetic component given by
\begin{equation}
\left[\frac{d \sigma (E_\nu, T_e)}{dT_e}\right]_{\mathrm{EM}} = \frac{\pi \alpha_{\mathrm{EM}}^2 \mu_{\nu,\mathrm{eff}}^2 }{m_e^2} \left(\frac{1}{T_e} - \frac{1}{E_\nu} \right) \, ,
\label{eq:xsec_EM}
\end{equation}
where $\alpha_{\mathrm{EM}}$ denotes the fine structure constant.\\[-.3cm]

The XENON1T collaboration has suggested a neutrino magnetic moment of the order of $2 \times 10^{-11}\mu_B$ as a way to account for the detected excess.
Since it is generally expected that neutrinos are Majorana fermions~\cite{Schechter:1980gr}~\footnote{The issue can only be
  settled experimentally by the detection of neutrinoless double beta decay~\cite{Schechter:1981bd}.}, it is therefore interesting to
take up their suggestion within the general framework of non-zero Majorana neutrino TMMs.
  The general parameterization in terms of the TMM matrix allows us to make a direct comparison of experimental
  results coming from different neutrino sources, e.g. solar, reactor
  or accelerator neutrino sources~\cite{Grimus:2002vb,Canas:2015yoa,Kosmas:2015sqa,Miranda:2019wdy}.
  Within this formalism, the effective neutrino magnetic moment is expressed as
\begin{equation}
\mu_{\nu,\text{eff}}^2 = \tilde{\mathfrak{a}}_{-}^\dagger \tilde{\lambda}^\dagger \tilde{\lambda} \tilde{\mathfrak{a}}_{-} + \tilde{\mathfrak{a}}_{+}^\dagger \tilde{\lambda} \tilde{\lambda}^\dagger \tilde{\mathfrak{a}}_{+} \, , 
\label{eq:TMM-mass}
\end{equation} 
where $\mathfrak{a}_{+}$ and $\mathfrak{a}_{-}$ are the $3-$vector amplitudes of positive and negative helicity states~\cite{Grimus:2000tq}, and the TMM matrix in the mass basis is given as
\begin{equation}
\tilde{\lambda} = \left( \begin{array}{ccc}
0 & \Lambda_3 & - \Lambda_2 \\
- \Lambda_3 &  0 & \Lambda_1 \\
\Lambda_2 & - \Lambda_1 & 0
\end{array} \right) \, . 
\label{NMM:matrix}
\end{equation}

In the general case, the effective neutrino magnetic moment takes into account neutrino propagation effects, and can be written as~\cite{Beacom:1999wx}
\begin{equation}
\mu_{\nu,\text{eff}}^2 (L, E_\nu) = \sum_j \Big \vert \sum_i U^\ast_{\alpha i} e^{-i\, \Delta m^2_{ij} L /2 E_\nu} \tilde{\lambda}_{ij} \Big \vert^2 \, ,
\label{NMM-observable}
\end{equation}
where $U_{\alpha i}$ are the elements of the lepton mixing matrix, $\Delta m^2_{ij}$ denote the neutrino oscillation mass splittings, $L$ is the distance travelled by the neutrinos, and $\tilde\lambda_{ij}$ are the elements of the TMM matrix in the mass basis, given at Eq.~(\ref{NMM:matrix}).
One sees how the effective neutrino magnetic moment at the experimental site will depend not only on the TMM matrix, but also on the mixing parameters which take into account the oscillation effects from the neutrino source to the detector. Therefore, one should be careful when comparing results on effective magnetic moments obtained at different experimental setups. In fact, to avoid potential confusion while performing such comparisons, we strongly recommend the use of the general TMM formalism.\\[-.3cm]

Under this hypothesis, the excess recoil events reported by XENON1T would be due to a neutrino TMM interaction of the solar
  neutrinos in the detector. The effective neutrino magnetic moment in this case is given as~\cite{Grimus:2002vb,Canas:2015yoa}
\begin{equation}\label{eq:nmm_sun}
\mu_{\nu,\,\text{sol}}^{2} = |\mathbf{\Lambda}|^{2} -
  c^{2}_{13}|\Lambda_{2}|^{2} + (c^{2}_{13}-1)|\Lambda_{3}|^{2} +
  c^{2}_{13}P^{2\nu}_{e1}(|\Lambda_{2}|^{2}-|\Lambda_{1}|^{2})\, ,
\end{equation}
with $c_{13} = \cos\theta_{13}$. For $P_{e1}^{2\nu}$, we consider the average value of this probability for solar $pp$ neutrinos which, within the 1$\sigma$ range obtained in global fits to neutrino data~\cite{deSalas:2020pgw}, takes the value $P_{e1}^{2\nu} = 0.667 \pm 0.017$. 
As mentioned before, the $\Lambda_i$ correspond to the components of the Majorana neutrino TMM matrix in the mass basis, and $\mathbf{\Lambda} = |\Lambda_1|^2 + |\Lambda_2|^2 + |\Lambda_3|^2$. Note that, in the particular case of solar neutrinos, the effective magnetic moment $\mu_{\nu,\,\text{sol}}^{2}$  is  independent of any CP violation phase~\cite{Canas:2015yoa}. 


\section{XENON1T low-recoil signal from neutrino TMM }
\label{sec:TMM-sol}

The $1/T_e$ term of the EM cross section in Eq.~(\ref{eq:xsec_EM}) leads to an enhancement of the predicted signal at low recoil energies.
As a result, the inclusion of EM neutrino interactions can lead to a low-energy bump in the measured spectrum,
motivating us to perform a sensitivity analysis based on the binned $\chi^2$ function
\begin{equation}
\chi^2 = \sum_{i=\mathrm{bins}}\frac{1}{\sigma_i^2}\left(\frac{dN^i_\mathrm{obs}}{dT_{rec}} - \frac{dN^i_\mathrm{th}}{dT_{rec}} \right)^2 \, ,
\end{equation}
where the index $i$ runs over the $i$-th bin of the observed XENON1T signal i.e. $dN^i_\mathrm{obs}/dT_{rec}$, with statistical uncertainty  $\sigma_i$. Our calculated number of events $dN^i_\mathrm{th}/dT_{rec}$ includes the  background $B_0$ reported by XENON1T, containing
 the solar neutrino background due to SM weak interactions as described in Eqs.~(\ref{eq:dNdT_SM})-(\ref{eq:xsec_SM2}), 
as well as the contribution due to the presence  of a non-zero neutrino magnetic moment given by Eq.~(\ref{eq:xsec_EM}).

\begin{figure}[t]
\includegraphics[width= \textwidth]{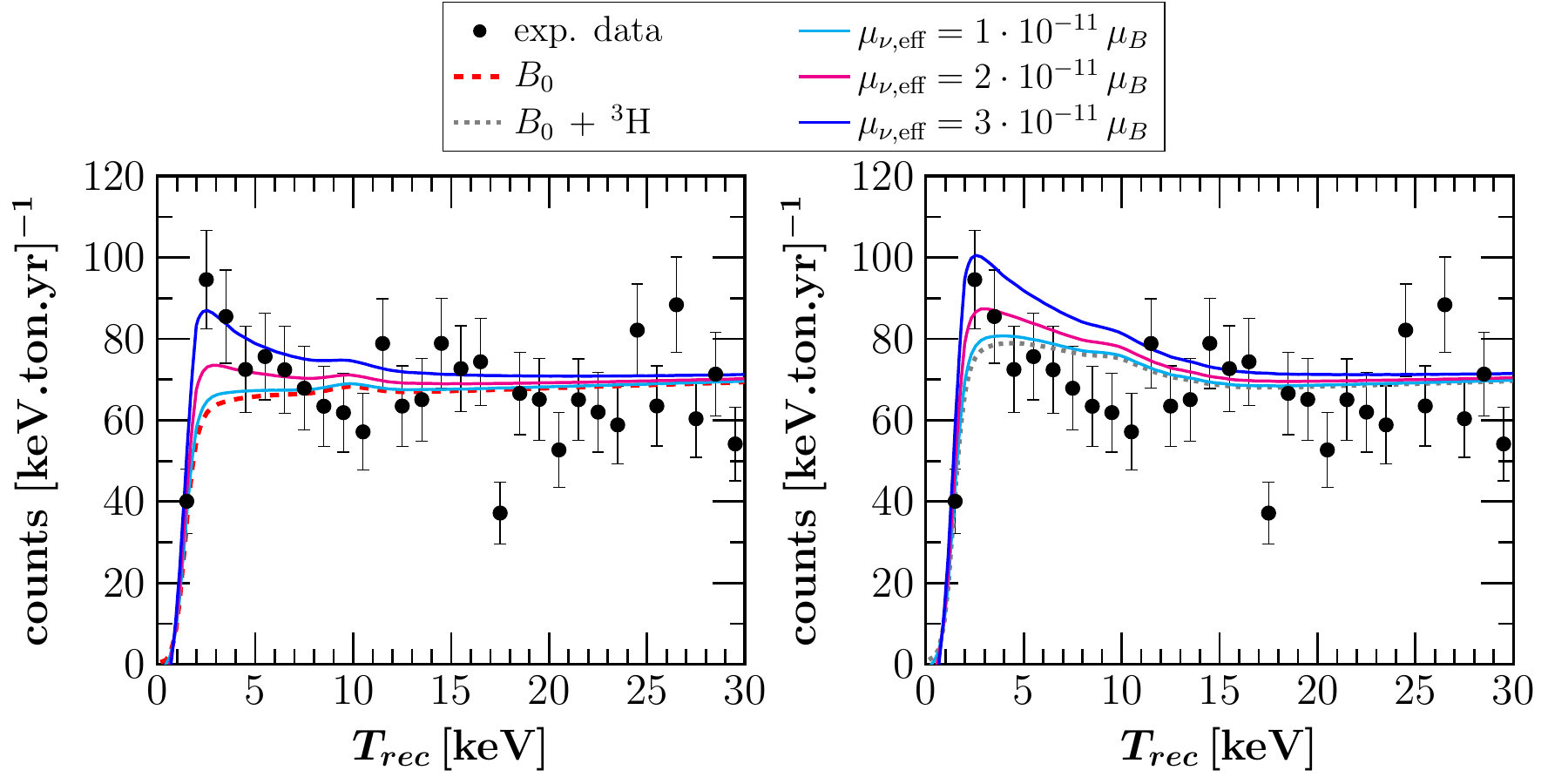}
\caption{ Left: comparing the XENON1T data with expected count rates from background only, and from the indicated values of the effective neutrino magnetic moment.
  Right: Same as left panel, but including possible tritium contamination. More details in text.}
\label{fig:diffrate}
\end{figure}

In the left panel of Fig.~\ref{fig:diffrate}, we compare the effect of EM neutrino interactions in the XENON1T detector, assuming the reported background $B_0$ only. 
Indeed, one can see a clear enhancement of the predicted EM signal at low recoil energies, as expected from the  $1/T_e$ dependence of the EM cross section.
We illustrate the effect by considering $\mu_{\nu,\text{eff}}=1-3\times10^{-11}\mu_B$, which corresponds to the range extracted by the XENON1T collaboration~\cite{Aprile:2020tmw}.\\[-.3cm]

In their effort to describe the excess of low-recoil events, the XENON1T collaboration has examined the possibility of additional $^{3}$H contamination.
Although it can produce an excess of low-recoil events, this background can not fully explain the observed data.
On the other hand, it seems fair to say that such $^{3}$H background is not fully understood.
It seems therefore worth exploring the impact of the total background, i.e.  $B_0+^{3}$H, in the presence of an effective neutrino magnetic moment.
This is shown in Fig.~\ref{fig:diffrate}. 
\begin{figure}[b]
\includegraphics[width =\textwidth]{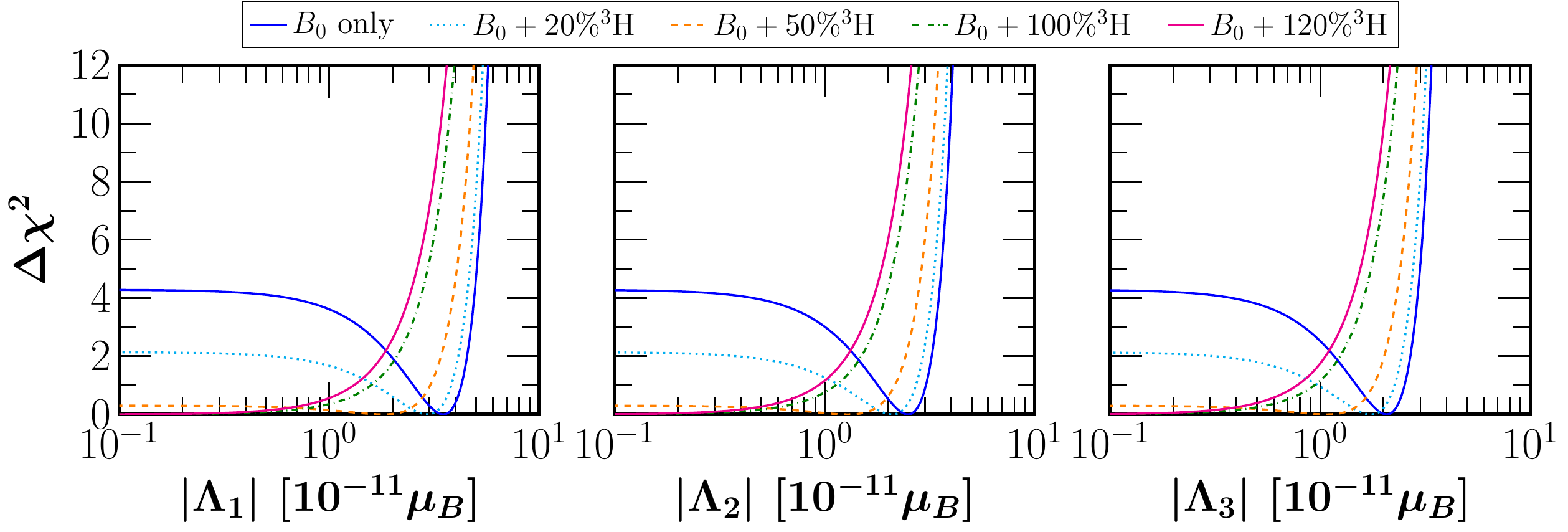}
\caption{XENON1T constraints on the effective neutrino magnetic moment for various choices of tritium background. 
}
\label{fig:chi2}
\end{figure}
One sees from the right panel of Fig.~\ref{fig:diffrate} that, while the $^{3}$H background is not, by itself, capable of accounting for the low-recoil excess, in the presence of moderate values of
$\mu_{\nu,\text{eff}}$, it does.
Indeed, we analyze how different fractions of the $^{3}$H component can affect the resulting constraints on $\mu_{\nu,\text{eff}}$.

In addition to the case without a $^3$H background, we have also explored the effect of a non-zero $^{3}$H contamination, varying it from 0 (no $^{3}$H) to 100\% (total $^{3}$H background considered in~\cite{Aprile:2020tmw}), as well as an extreme case with a 120\% $^3$H
contamination~\footnote{This may well account for any other unknown background in addition to tritium~\cite{Bhattacherjee:2020qmv}.}. 
The resulting $\chi^2$ profiles for the elements of the TMM matrix $\Lambda_i$ taken one at a time are illustrated in Fig.~\ref{fig:chi2}.
As expected, one finds that, for larger fractions of the $^{3}$H background, the corresponding values of the neutrino magnetic moment decrease, and could lie in the sub-$10^{-11}\, \mu_B$ regime.  


\section{Comparing with other limits}
\label{sec:other-experiments}

\begin{figure*}[t]
\includegraphics[width=\textwidth]{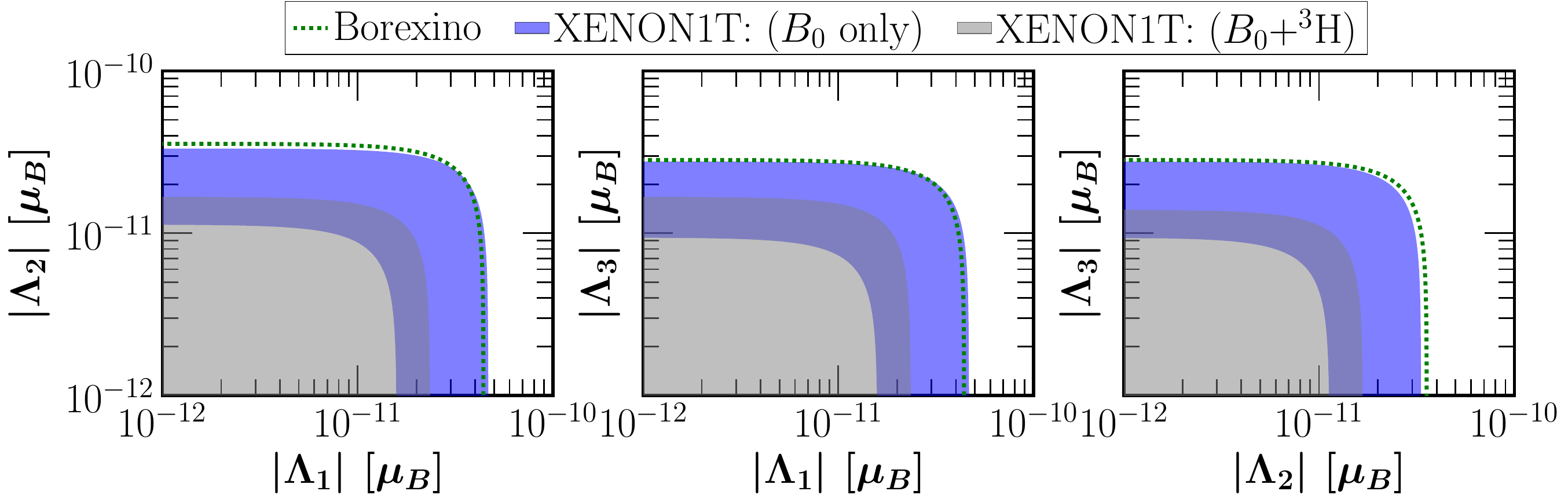}
\caption{ 
  Regions in the $\left \vert \Lambda_i \right \vert - \left \vert \Lambda_j \right \vert$ plane allowed by the XENON1T data at 90\% C.L.
 The outer blue band assumes no tritium background, in constrast to the inner grey region. Note that these bands overlap.   The results obtained by Borexino (dotted line) are shown for comparison.}
\label{fig:Li_vs_Lj_Borexino}
\end{figure*}

Besides the one-dimensional constraints presented in Fig.~\ref{fig:chi2}, here we show in Fig.~\ref{fig:Li_vs_Lj_Borexino} the 90\% C.L. regions in the two-dimensional planes $|\Lambda_i|$-$|\Lambda_j|$ allowed by the recent XENON1T data. The blue bands have been derived by considering only the background $B_0$, while the grey regions correspond to the analysis with $B_0$ + $^3$H background. 
 These limits on the TMMs are obtained by taking two parameters at a time, and assuming a vanishing value for the third (undisplayed) $\left \vert \Lambda_k \right \vert$.
  The Borexino limits~\cite{Borexino:2017fbd} on the TMMs as derived in Ref.~\cite{Miranda:2019wdy} are also indicated by the dotted green lines. 
  One sees from the figure that the TMM values required by our proposal are competitive with the current limit reported by the Borexino collaboration.
  Indeed, the current Borexino limit lies close to the magnitude indicated by the explanation of the low-recoil data, so that future measurements should be able to explore it with accuracy.
\\[-.3cm]
%

In Table \ref{table:NMM} we  also give the 90\% C.L. Borexino constraints on the TMM matrix elements $|\Lambda_i|$ as obtained in \cite{Miranda:2019wdy}, as well as the allowed ranges and bounds derived from XENON1T data for different assumptions on the background. All results are fully consistent. \\[-.3cm]
\begin{table}[!t]
\begin{tabular}{l|ccc}
\hline
  & \, $|\Lambda_1|\,[10^{-11} \mu_B]$ \, & \, $|\Lambda_2|\,[10^{-11} \mu_B]$ \, & \, $|\Lambda_3|\,[10^{-11} \mu_B]$ \, \\ 
\hline
Borexino~\cite{Miranda:2019wdy} & < 4.4         & < 3.6    & < 2.8         \\
XENON1T  ($B_0$ only) & (1.5--4.7)    & (1.1--3.3)    & (0.9--2.8)     \\ 
XENON1T  ($B_0$ + $^3$H) & < 2.4 & < 1.7 & < 1.4 \\
\hline
\end{tabular}
\caption{ 90\% C.L. limits and allowed ranges for the TMM matrix elements $|\Lambda_i|$ obtained from Borexino and the recent XENON1T data under different background assumptions. }
\label{table:NMM}
\end{table}

Beyond the Borexino limit on the effective neutrino magnetic
  moment, there are laboratory constraints coming from reactor
  neutrinos such as the GEMMA~\cite{Beda:2012zz},
  TEXONO~\cite{Deniz:2009mu}, and MUNU
  experiments~\cite{Daraktchieva:2005kn}, from
  LSND/LAMPF~\cite{Auerbach:2001wg,Allen:1992qe}, as well as by analyses of coherent elastic neutrino nucleus scattering (CE$\nu$NS) data~\cite{Miranda:2019wdy}.
  These limits are less stringent than the Borexino bound and, more important, constrain a quite different effective neutrino magnetic moment and, therefore, a different set of parameters that also include non-trivial CP phases~\cite{Canas:2015yoa,Miranda:2019wdy}. Hence, a direct comparison with such results is not as interesting and straightforward as it is with Borexino.
In any case, one clearly sees that the magnetic moment strengths required to explain the XENON1T excess agree with current experimental bounds.\\[-.3cm]

Astrophysics, however, places more stringent limits~\cite{Raffelt:1990pj,Heger:2008er}.
However, as we have seen, due to the currently unknown level of tritium contamination, one can account for the observed low-energy
recoil excess with smaller values of the effective neutrino magnetic moment, potentially avoiding tension with astrophysical limits. \\[-.3cm]

Nevertheless, the presence of a finite neutrino magnetic moment could also affect the propagation of solar neutrinos beyond the oscillation mechanism~\cite{Akhmedov:1988uk}.
This happens, for example, as neutrinos cross the convective zone of the Sun, which could host large magnetic fields.
In this case, there is a rich interplay between the pure oscillation mechanism and the effect of transition neutrino magnetic moment,
leading to an effective anti-neutrino component in the solar neutrino flux.
Assuming that the convective zone harbors random magnetic fields, this leads to an enhanced expected solar electron antineutrino flux~\cite{Miranda:2003yh,Miranda:2004nz}.
  In fact, assuming the same turbulent magnetic field model employed in~\cite{Miranda:2004nz}, the bounds implied by Super-Kamiokande and KamLAND would be $5\times 10^{-12} \mu_B$.
  However, this result depends on the choice of scaling law for the turbulent kinetic spectrum, and other choices could give somewhat weaker bounds, see~\cite{Miranda:2004nz} 
  for a more detailed discussion. In any case, these considerations open up a way of probing the constraints extracted in our proposal.

\section{ conclusions}

  Many possible explanations have been suggested for the recent XENON1T collaboration event excess at few-keV recoil,
  see for instance Refs.~\cite{AristizabalSierra:2020edu,Boehm:2020ltd, Lindner:2020kko,DiLuzio:2020jjp, Gao:2020wer}.
Here we have examined the possibility that the excess could be due to the scattering of solar neutrinos carrying a small neutrino transition magnetic moment.
We have found that the observed event excess is consistent with the neutrino TMM interpretation and its strength is in agreement with current experiments.
Future low energy experiments using artificial neutrino sources~\cite{Coloma:2020voz,Link:2019pbm} or coherent elastic neutrino nucleus scattering~\cite{Kosmas:2015sqa,Miranda:2019wdy} could also help confirming or ruling out our proposal.
An important role is played by the presence of the tritium background, so far not fully understood.
The larger that contamination is, the smaller the magnetic moment strengths required to explain the excess, thus relaxing possible conflict with astrophysics.

\begin{acknowledgments}

  The authors are grateful to D. Aristizabal, V. De Romeri and L. Flores from fruitful discussions.
This work is supported by the Spanish grants FPA2017-85216-P (AEI/FEDER, UE), PROMETEO/2018/165 (Generalitat Valenciana) and the
Spanish Red Consolider MultiDark FPA2017-90566-REDC, and by CONACYT-Mexico under grant A1-S-23238. OGM has been supported by SNI
(Sistema Nacional de Investigadores).
The work of DKP is co-financed by Greece and the European Union (European Social Fund- ESF) through
the Operational Programme <<Human Resources Development, Education and Lifelong Learning>> in the context of the project ``Reinforcement of
Postdoctoral Researchers - 2nd Cycle" (MIS-5033021), implemented by the State Scholarships Foundation (IKY).
MT acknowledges financial support from MINECO through the Ram\'{o}n y Cajal contract RYC-2013-12438.

\end{acknowledgments}

\black
\bibliography{bibliography} 
\bibliographystyle{utphys}


\end{document}